# Three-dimensional atomically-resolved analytical imaging with a field ion microscope

Shyam Katnagallu[a], Felipe F. Morgado[a], Isabelle Mouton[a,$], Baptiste Gault[a,b], Leigh T. Stephenson[a,*]

[a]*Department of Metal physics and alloy design, Max Planck Institut für Eisenforschung GmbH, Düsseldorf 40237, Germany*

[b]*Department of Materials, Royal School of Mines, Imperial College London, London, SW7 2AZ, UK*

[$] *now at CEA Saclay Des/-Service de Recherches de Métallurgie Appliquée, Gif-sur-Yvette, France.*

**Abstract**

*Atom probe tomography (APT) helps elucidate the link between the nanoscale chemical variations and physical properties, but it has limited structural resolution. Field ion microscopy (FIM), a predecessor technique to APT, is capable of attaining atomic resolution along certain sets of crystallographic planes albeit at the expense of elemental identification. We demonstrate how two commercially-available atom probe instruments, one with a straight flight path and one fitted with a reflectron-lens, can be used to acquire time-of-flight mass spectrometry data concomitant with a FIM experiment. We outline various experimental protocols making use of temporal and spatial correlations to best discriminate field evaporated signals from the large field ionised background signal, demonstrating an unsophisticated yet efficient data mining strategy to provide this discrimination. We discuss the remaining experimental challenges that need be addressed, notably concerned with accurate detection and identification of individual field evaporated ions contained within the high field ionised flux that contributes to a FIM image. Our hybrid experimental approach can, in principle, exhibit true atomic resolution with elemental discrimination capabilities, neither of which atom probe nor field ion microscopy can individually fully deliver – thereby making this new approach, here broadly termed analytical field ion microscope (aFIM), unique.*

*KEYWORDS: analytical field ion microscopy, time-of-flight mass spectroscopy, atom probe tomography, atomic resolution, spectral decomposition*

## 1. Introduction

Engineering properties are fundamentally a consequence of a material's atomic architecture and in this regard modern high-resolution microscopes have provided invaluable insights that have led to significant advances in materials' design and production. Extracting detailed structural and chemical information from a single microscopy technique like atom probe tomography (APT) or transmission electron microscopy (TEM) is desirable though especially challenging. Achieving an

---

atomically and chemically resolved three dimensionally significant material makeup with a relative ease is as yet an unrealised dream.

Field ion microscopy (FIM) realised the first images of individual atoms (Müller & Bahadur, 1956). The field ionisation of imaging gas atoms and the field evaporation of surface atoms were enabled through applying a sufficiently high standing voltage, and they can be controlled in combination with voltage pulsing (Müller et al., 1968; Müller & Bahadur, 1956) or laser pulsing (Cerezo et al., 2007). By finely controlling the field evaporation rate, a resulting 3DFIM image stack was shown to reveal crystalline lattice complete with defects of a pure iron sample (Vurpillot et al., 2007) or pure W (Dagan et al., 2017). Further studies played on this capability for an array of materials (Akré et al., 2009; Danoix et al., 2012; Cazottes et al., 2012; Koelling et al., 2013). Meticulous attention has lately been devoted to the 3DFIM reconstruction, both in developing automated protocols (Katnagallu, Gault, et al., 2018; Dagan et al., 2017) and describing the dynamic aberrations that affect field ion microscopy imaging (Katnagallu, Dagan, et al., 2018).

Atom probe tomography (APT) developed from FIM. FIM designs (Ingham et al., 1954) evolved into atom probe - field ion microscopes (AP-FIM) (Kellogg & Tsong, 1980; Müller et al., 1968). These instruments provided both time-of-flight mass spectrometry (TOF-MS) and three-dimensional atomic reconstructions with sub-nanometre spatial resolution (Miller & Forbes, 2014; Seidman, 2007; Miller et al., 2012; Kelly & Miller, 2007). APFIM designs could be switched between a field ionisation mode to an analytical mode, based on the atom probe mass spectrometer. In FIM, micrographs were acquired on a dedicated, movable screen comprising an assembly of imaging micro-channel plates and a phosphor anode. Pulsed field-evaporation for TOF-MS was done through an atom probe hole into a long differentially-pumped drift tube and the time-of-flights of any field evaporated ions were measured on a electron multiplier detector. Such early AP-FIM experiments achieved good TOF-MS with a low background level (Müller, 1971).

The design of AP has evolved over several decades to include position-sensitive detectors to enable three-dimensional reconstructions, pulsed-laser sources to analyse poor electrical conductors (Hono et al., 2011; Kellogg & Tsong, 1980), and, sometimes, an electrostatic lensing systems (Camus & Melmed, 1990; Panayi et al., 2006; Bémont et al., 2003; Sakai & Sakurai, 1984), usually to improve the mass-to-charge discrimination capabilities. Switching modes was not rapidly done and now, with modern atom probe laboratories aiming for high throughput, careful maintenance of ultrahigh vacuum conditions is the usual practice, mostly to avoid unwanted, difficult-to-identify peaks in the mass-to-charge spectrum and higher than necessary levels of background. Modern instruments employ wide field-of-view, position-sensitive detectors with high speed electronics that enabled the analysis of larger volumes (Kelly et al., 2004). The early versions of the wide field-of-view commercial atom probes often included electronic FIM (eFIM) or digital FIM capability, along with a set of imaging gas bottles and a FIM mixing chamber. The latest generation of instrument is only rarely equipped with such modules. eFIM is a digital image recalculation of the field ionised impacts on the APT detector itself. The high flux of incoming imaging gas ions, i.e. several thousands of ions per second for each imaged surface atom, is orders of magnitude higher than what is usually encountered in APT and pushes the detector towards its intrinsic limits.

Here, we discuss a hybrid FIM-APT experiments using our straight flight path local electrode atom probe (Cameca LEAP-5000XS), as well as an atom probe equipped with a reflectron (Cameca LEAP 3000X HR), both fitted with a gas mixing chamber hence amenable to eFIM – as summarised in Figure 1(a). Figure 1(b) demonstrates the imaging capabilities of eFIM for tungsten.

Due to a lack of pulsing capability in the eFIM software module, it was not possible to achieve a controlled field evaporation while also maintaining reasonable imaging conditions. Imaging field evaporating tungsten requires pulsing capability, since the evaporation field of tungsten is in the range of 55Vnm$^{-1}$, and at such fields, the field ionisation of He proceeds too far from the specimen's surface to allow for atomically-resolved imaging. High-voltage pulsed FIM has long been reported to enable controlled field evaporation of the specimen's surface atoms during FIM imaging (Kellogg & Tsong, 1980). Laser-pulsed FIM experiments were also previously discussed in terms of the individual space-time coordinate of individual field-ionised detection events (Silaeva et al., 2014; Vurpillot et al., 2009; Kim & Owari, 2018).

To move beyond the aforementioned limitations of the eFIM module, we simply used the standard APT software module to provide automatic and periodic voltage pulsing in the manner described below. In doing so, we aspired to combine the atomic resolution of eFIM imaging and the mass-to-charge discerning abilities of an APT acquisition under gas pressures enabling imaging, here describing this technique as analytical field ion microscopy or aFIM. We treated pressure as a new experimental parameter spanning the continuum between the ultrahigh vacuum of APT and the high vacuum of FIM. The scope of this manuscript is to describe the set of preliminary experiments that demonstrate the acquisition and analysis of mass-to-charge information and its connection to meaningful FIM images. This in parallel to our other study, wherein chemical information in a binary alloy was inferred from correlating image brightness with atomic identity via density functional theory simulations (Katnagallu et al., 2019). We demonstrate this for pure tungsten, acquiring mass-to-charge spectra with admittedly a massive imaging gas background. This background can be significantly filtered from correlated detection events representing the co-evaporation of metal and the adsorbed gas atoms. We also consider the utility of applying sophisticated data mining and image analysis techniques to further discriminate the pulse-correlated evaporating ions and the continuous gas ion signal. Finally, we suggest experimental protocols to maintain good operational conditions for further aFIM experiments and later APT experiments.

## 2. Methods

### 2.1. Materials

To establish proof-of-concept aFIM, pure tungsten needles were prepared from a drawn wire most often exhibiting a <110> z-axis orientation, by using electrochemical polishing at 5-8 VAC in a 5% molar NaOH solution. We also employed as specimens pre-sharpened doped Si-microtips that are commercially available by Cameca Instruments (Kelly & Miller, 2007).

### 2.2. Operating FIM on the LEAP systems

The Cameca Local Electrode Atom Probe (LEAP) 5000XS is shown schematically in **Error! Reference source not found.**(a). It is equipped with a FIM mixing chamber, with three 99.999% pure gases (He, Ne, Ar). To assess how a reflectron could affect the results, we also performed similar experiments on a Cameca LEAP 3000X HR (now decommissioned), also equipped with FIM capabilities. The acquisition software for pulsed APT and eFIM modes are Cameca's own proprietary software, so-called DAVIS. It was slightly different for each machine, yet our experiments could be equally performed on either of the instruments, and have been performed since the latest upgrade of the acquisition software, now part of AP-Suite, on the LEAP 5000 XS.

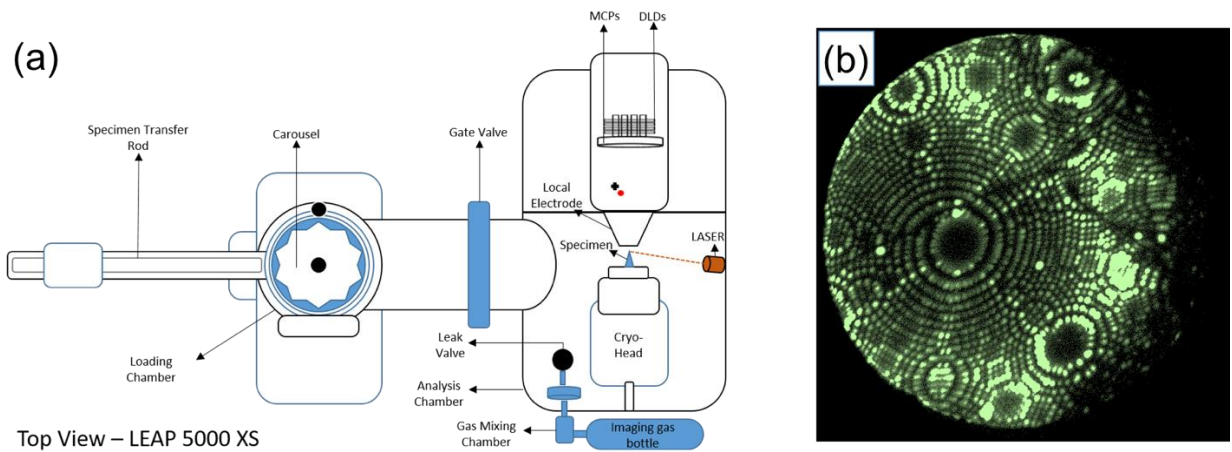

*Figure 1: (a) A schematic showing the top view of the LEAP 5000XS atom probe instrument which is additionally equipped with imaging gas bottles for field ion microscopy. (b) An example of the electronic FIM, or "eFIM", that is achievable with the Cameca software on the LEAP 5000XS, this being the surface of an electropolished drawn tungsten wire. The surface presented exhibits the interfaces of multiple grains.*

We implemented standard practice to establish a pure gas in the analysis chamber, flushing and evacuating the gas mixing chamber twice with a pure gas, and refilling it to approximately 1 Torr (using this measure of pressure only because it is common across modern LEAP machines – with 1 Torr being approx. 133.3 Pa). For FIM imaging, we set the specimen temperature to 50K and, with a manual leak valve, we then admitted the imaging gas into the instrument's analysis chamber to the desired imaging pressure, usually in the range of $1 \times 10^{-7}$ Torr. The Cameca eFIM software module only provides parameters to emulate traditional phosphor screen imaging, i.e. exposure time, life-time and decay of events, and the digital imaging parameters of brightness, contrast and gamma. Other than that, this module provides manual control over the applied standing voltage bias, which allows the voltage to be manually ramped up to the best-imaging voltages (BIV) and the field evaporation voltages. The eFIM module appears to store the data pertaining to individual hits, the vast majority of which during a normal FIM experiment would be the imaging gas ions, all it really provides access to is the possibility to record and play videos of the recalculated FIM imaging sequence. The output from this operating mode is not suitable for accurate 3DFIM reconstructions. Figure 1(b) gives an example of the eFIM imaging available on the LEAP 5000XS. In such a case, there can be no simple time-of-flight discrimination as there was no timed pulse (either voltage or laser pulse). The delay-line detector (DLD) channel information establishes only the intensity of the gas ion flux at different projected positions. We simply operated the standard APT software in acquisition mode. While both high-voltage and laser pulsing could be used, we focused on voltage pulsing in this work.

## 3. Results and Discussion

*3.1. Pulsed FIM in the LEAP*

Voltage pulsing was defined and controlled primarily through the pulse fraction: a specimen was held at a positive standing voltage and the field experienced by the specimen's surface atoms was then enhanced by negative pulses upon the opposing local electrode, these pulses being a constant fraction of the standing voltage. Implementing pulsed FIM upon the LEAP 5000XS had no good standard practice, so we initially approached this cautiously. We were unsure whether the detector and supporting electronics could make much sense of the expected large flux. Unlike eFIM, this pulsed FIM-in-APT mode, i.e. aFIM, records and stores every detector event, including detector coordinates and the time-of-flight relative to the time of the last pulse.

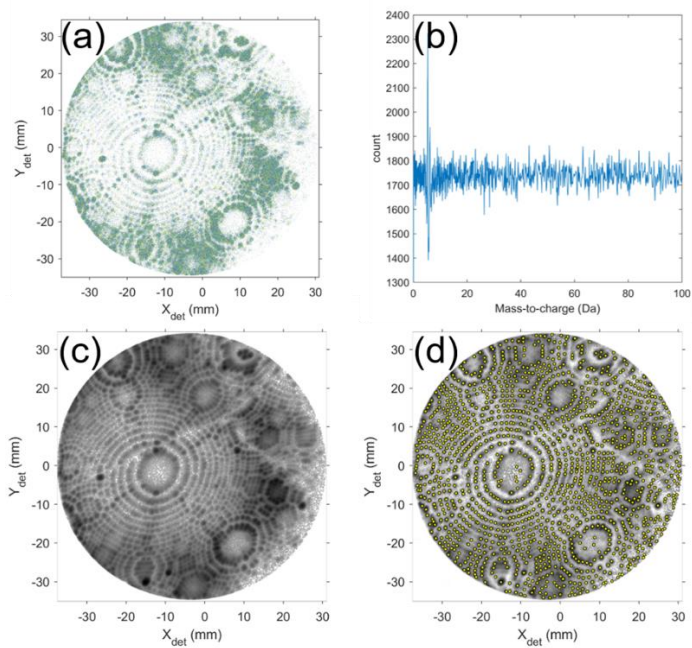

Figure 2. *Experiment on a specimen of drawn tungsten wire FIM in APT mode using $8 \times 10^{-8}$ Torr He and a best-imaging voltage of 5kV and a 500V pulse and a 250kHz pulse frequency. (a) A subset of $10^5$ individual field ionised events recorded on the detector within a narrow time interval over several thousand voltage pulses (<1% pulse fraction). These detector coordinates are made accessible through the EPOS file format. The same surface was imaged afterwards using standard e-FIM (Figure 1) (b) The mass-to-charge spectrum calculated from a TOF spectrum roughly corrected with Pythagoras' theorem for longer ion flights away from the detector centre with quadratically-growing bin widths corresponding to linearly-spaced bins in the TOF space. (c) A 2048-by-2048 FIM image compiled from $6 \times 10^6$ points from the same dataset as (a). The image is generated by first using a two dimensional kernel density estimation (2DKDE) and then logarithmically scaling the image intensity. (d) Local contrast was then enhanced and maximum intensities levelled using Contrast Limited Adaptive Histogram Equalisation (CLAHE) and then, by filtering out all signal below 70% and using a peak-finding method that convolutes the image with a 25-by-25 Gaussian kernel ($\sigma = 5$ bins), the individual surface atoms could all be identified with few apparent false positives.*

Figure 2(a) demonstrates the possibility of performing FIM-type imaging. We used voltage pulsing and used a 1% pulse fraction and introduced $8 \times 10^{-8}$ Torr He gas. It was thought at this stage that the individual impact coordinates were mostly of field desorbed and field ionised helium ions. Figure 2(b) shows a mass-to-charge spectrum attained through a naïve application of the IVAS time-of-flight corrections which mostly evidenced a smooth background and a small feature near where we would expect a 4-Da helium peak. This is later discussed herein. Such automatic time-of-flight corrections are difficult without a prominent signal and a more sophisticated and tailored approach would be required to attain reasonable mass-to-charge spectra such as described in (Caplins et al., 2020). As commonly seen in conventional APT experiments, the uncorrelated background signal has a uniform distribution in the time-of-flight space but has an inverse square root distribution when transformed to the mass-to-charge space. We took advantage of this in Figure 2(b), giving the mass-to-charge spectrum a uniform background by calculating the histogram with quadratically-growing bin widths (equivalent to equally-spaced bins in the TOF space). This gives clarity to any signal that emerges above the background, particularly in the low Dalton range.

As seen in Figure 2(a), some surface atoms were brightly imaged by FIM whereas other surface atoms could barely form a discernible image, due to a lack of collected signals. One approach would have been to increase the pressure to improve FIM imaging. Instead, we increased the exposure time over which we integrated the detector coordinate data. Like atom probe, aFIM is regularly pulsed so this time could be measured in pulses or in seconds. Figure 2(c) shows an image constructed out of an exposure with six million ions, 60 times the data of Figure 2(a). Rather than a simple 2D histogram, which can contain considerable noise, we instead applied an adaptive 2D kernel density estimation (2D-KDE) routine (Botev et al., 2010), and this provided a smooth image that can better display individual atoms. Even processed like this, the 2D-KDE image has a high dynamic range, so we used a logarithmic scaling for the greyscale image intensity. Other tone-mapping methods could be used to visualise and reduce the data (Katnagallu, Gault, et al., 2018). Figure 2(d) demonstrates an application of convolutional peak finding using a 2D Gaussian function as a model of the surface atoms' images. Enhancing local contrast and equalising intensities via the CLAHE algorithm ("contrast limited adaptive histogram equalization" (Zuiderveld, 1994)) was also trialled to attain reasonable results out of the peak finding. In this way, surface atoms were automatically identified in a FIM snapshot. This step is a prerequisite for performing three-dimensional reconstructions following the protocols introduced in Ref. (Dagan et al., 2017; Katnagallu, Dagan, et al., 2018) for instance.

*3.2. Analytical FIM principle*

The principle of aFIM is the same as for pulsed FIM, except we obtain detector coordinates and a mass-to-charge spectrum similar to an APT experiment. Using this mode was intended to exploit two processes:

- The first process occurs between pulses at a relatively low standing voltage when the imaging gas atoms are field ionised and, when detected, these can be accumulated into a FIM image. Ideally, the standing voltage is at the specimen's best imaging voltage (BIV). Events are ions that have no correlation with a timed pulse and so contribute to a uniform time-of-flight spectrum and the smooth background commonly seen in APT experiments.
- The second process occurs during the short voltage pulse, or subsequent to the laser pulse, when the specimen's surface atoms are field evaporated. These ions can then be detected after the pulse and their time-of-flight registered. These ions collated together would resemble a raw APT dataset, together with a detector desorption histogram and a time-of-flight mass-to-charge spectrum.

We performed aFIM experiments with an objective to have two such easily-separable processes. Figure 3 illustrates a short survey of the some aFIM experiments when altering the voltage pulsing, showing both FIM imaging and the corresponding mass-to-charge spectra (the corrections for which are described below). Along the top margin of Figure 3, we summarise the voltage conditions used for the initial APT experiment and the four following aFIM experiments. The specimen was evaporated using APT to attain a clean desorption image (Fig 3(a)) and corresponding mass-to-charge spectrum (Fig. 3(b)). The following aFIM experiments were performed on a following day with a different local electrode and the 6kV BIV was previously selected from a preluding eFIM experiment not shown here, and for each consecutive aFIM experiment we increased the pulse fraction.

The LEAPs are equipped with an assembly of multi-channel plates (MCP) and a delay line detectors (DLD). In conventional experiments, ions are generally detected "one atom at a time", as trademarked by Cameca, or close enough to such conditions. This detector can efficiently discriminate multiple impacts emitted by a single pulse. In normal operation under APT mode, the detected events are sparse with respect to the pulsing. Our experience proved that we could produce similarly sparse evaporation for our hybrid aFIM experiments when using pressures within the $10^{-9}$ - $10^{-5}$ Torr range. For these experiments, the detector rarely ran over one detected ion per pulse on average (100% detection rate) but, being careful not to overtax the detector or the DLD signal processing, we often operated in a regime where the detection rate was less than 10% (0.1 ions/pulse).

The TOF corrections for these aFIM experiments were done manually as using the commercial software in this application was problematic. From the commercial software we obtained an extended-POS "EPOS" file (although it should be noted that this information will be accessible through the .APT file in the future (Reinhard et al., 2019)). This file yielded the data that is generally necessary for TOF corrections; the raw TOF data $T_0$, the detector coordinates $X_D$ and $Y_D$, standing voltage V, and the detection multiplicity and quiescent period (time since last detection, measured in pulses). For the aFIM experiments shown in Figure 3, the standing voltage was kept at 6kV and so a voltage TOF correction was not required to account for TOF variations as a function of voltage. Bowl corrections were required to account for the longer flight paths which occur in the wide field-of-view (Larson et al., 2013). These were performed on a case-by-case basis. For instance, the time-of-flight histograms in Figures 3(d/f) showed no discernible peaks that can be definitively assigned to field evaporation events; the ripples are caused by another effect and are dismissed largely as an artefact, later discussed, in any case, for those datasets, the Pythagoras' theorem was used to account for the longer flight paths moving away from the detector centre and this allowed for an approximate TOF correction.

For the other spectra (Fig. 3(h/j)), some field evaporation signal can be filtered from the field ionisation background by taking only multiple detection events or by the quiescent period (as in (Yao, 2016)). In the dataspace spanned by the raw TOF and the detector coordinates, the contrast of the field evaporation signal was enhanced by applying a threshold to a spatial statistic (e.g. kNN distance). This is particularly effective for multiple events as they tend to arrive in close proximity upon the detector (De Geuser et al., 2007). After filtering the background, we isolated the events we identified as desorbing $^{20}$Ne from the residual gas in the FI mixing chamber, and we fitted a quadratic 2D polynomial to express the raw TOF as a function of the detector coordinates. This was then used to transform the original TOF data $T_0$ to $T_1$. This process was repeated if necessary, focussing upon ions identified as W, and another quadratic 2D polynomial transforming corrected the time-of-flights from $T_1$ to $T_2$. We then constructed a $T_2$ time-of-flight spectrum, and fitted a 1D quadratic curve to relate the corrected TOF peaks with probable isotopic masses. Except for the filtering, this resembles a method already established for atom probe tomography (Sebastian et al., 2001).The final mass-to-charge spectra seen in Figures 3(h/j) as displayed using quadratically-growing bin widths as in Figure 2(b).

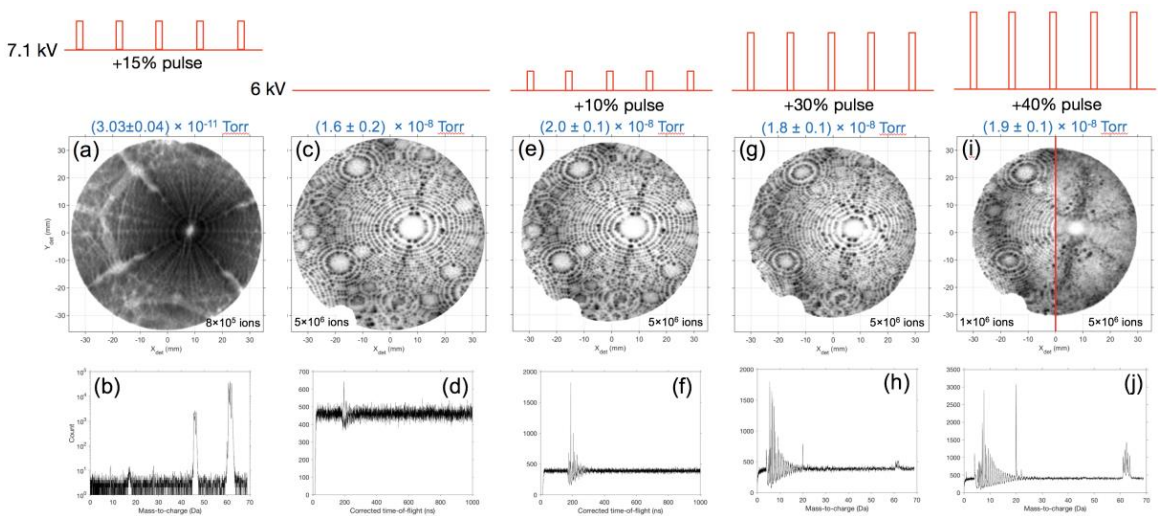

*Figure 3. A series of aFIM experiments varying voltage pulse fraction following a short atom probe experiment (acquired earlier with a different aperture) with a 15% pulse fraction and a 250 kHz pulse frequency. All experiments were performed at 50 K with approximately 2.0 x $10^{-8}$ Torr helium as imaging gas for aFIM. (a) & (b) APT field desorption histogram and mass-to-charge spectrum showing strong W3+ and W4+ peaks. (c/e/g/i) & (d/f/h/j) aFIM images and spectra for 0% (nominally), 10%, 30% and 40% pulse fractions respectively. Each aFIM image was 2048-by-2048 and processed as for the Figure 2(c). Note for (i): a red line divides two exposures, with a shorter exposure providing a cleaner FIM image.*

Figures 3(c) and 3(d) show the aFIM results for a nominal 0% pulse fraction; there it seems as though there is still a small pulse being delivered ≤ 1% pulse fraction. This was evidenced by the ripples in the Pythagorean-corrected TOF spectrum. Although these could not be assigned to any particular isotope, through looking at the later experiments it can be inferred that they soon occur after a pulse-correlated desorbed helium peak. Besides this feature, the spectrum of Fig. 3(d) displayed a uniform TOF background. Increasing the pulse fraction to 10% (Fig. 3(e/f)) resulted in an increase in the ripples following the expected desorbed helium peak (see discussion).

Increasing the pulse fraction to 30% (Fig. 3(g/h)) and 40% (Fig. 3(i/j)) had a number of effects on the mass-to-charge spectra. In Fig. 3(h/j), we saw pulse correlated events increasing with increasing the pulsed voltage. Peaks at 1-Da and 4-Da peaks emerged which were taken for desorbed monoatomic hydrogen and helium. At 30% pulse fraction, tungsten peaks (>60 Da) evidenced field evaporation but, surprisingly, there was also evidence of desorbing neon (20 Da) which may have been present either as a contaminant in the gas mixing chamber or residual from an earlier FIM experiment. Both W and Ne peaks were further pronounced above the background when using a 40% pulse fraction. Each of the FIM images was produced in the same manner as Figure 2(d). Figures 3(c/e/g) were obtained with 5-million ion exposures. The trend with increasing pulse fraction was seen as a smoothing of the image intensity around high index poles possessing a higher density of surface atoms. This was pronounced in Fig. 3(i) when we used a 40% pulse fraction. Here, terrace rings around the (110) pole showed little atomic definition. In this case, the evaporation was too fast for a long exposure and a shorter 1-million ion exposure produced a better quality image.

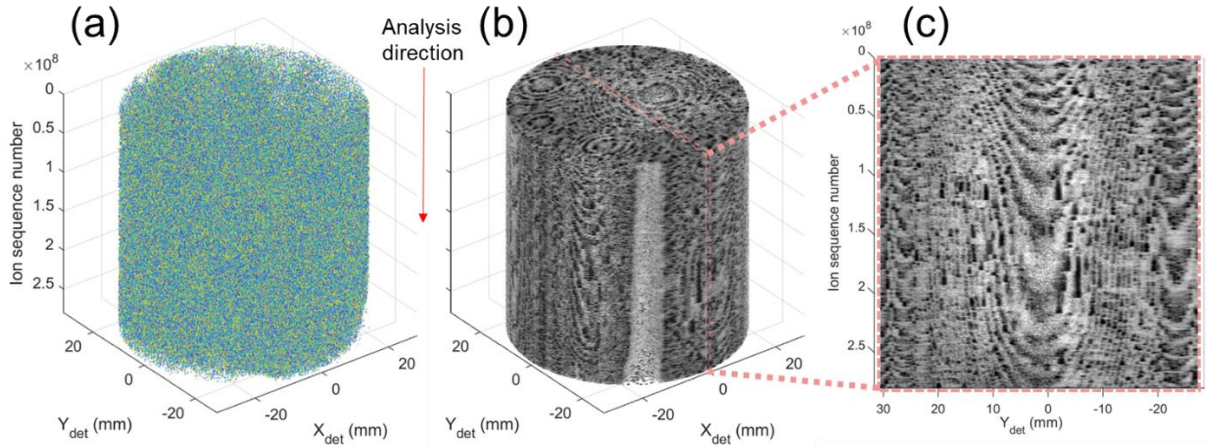

*Figure 4. 12-hour data acquisition of a tungsten tip held at 50 K and a 6-kV standing voltage with a 35% pulse fraction in the presence of 0.15 to 3.6 x $10^{-8}$ Torr of He. (a) A point cloud visualization of individually detected events in a space spanning the detector coordinate and ion sequence number. For clarity, only 1 million out of 280 million are shown and most of these events would be field ionized gas. The colour scale maps to the corrected mass-to-charge but is here only added to aid in visualization and to emphasize the data's pointillistic nature. (b) A stack of FIM images, each calculated from 500000 detected events by the image analysis methods performed in Figure 2, clearly showing the evolution of the specimen surface as field evaporation occurs. (c) A slice from the 3DFIM stack showing planes distorted by this simple ``reconstruction'' and, within the planes, surface atoms are clearly imaged. Prior to field evaporation, a surface atom leaves behind a plume of field ionized gas atoms.*

Just as for APT, aFIM experiment can be visualized as a three-dimensional point cloud. Figure 4(a) depicts the 280 million ions detected as part of an overnight experiment directly following the experiments of Figure 3. A constant 6-kV standing voltage was maintained with a 35% pulse fraction (2.1 kV pulse) and 250-kHz pulse repetition rate. The ions were visualised in a space spanning the detector coordinates ($X_{det}$, $Y_{det}$) and the ion sequence number ($z_i$). It would disingenuous to call this detector coordinates plus sequence number point cloud a reconstruction. Still, this approach has previously been utilized in cases where conventional APT data reconstructions cannot be informed by physical measurements or assumptions about specimen geometry (Rusitzka et al., 2018). The depth scaling on the $z_i$ can be variable if it was not kept at one-to-one as in Figure 4(a). Using the same image construction previously described for Figures 2 & 3, this ($X_{det}$, $Y_{det}$, $z_i$) point cloud was converted to a 3DFIM representation shown in Fig. 4(b), which is similar to a digitally-converted phosphor-screen image stack except in this case the images are produced by integrating 500000 events. The stacked images show the evolution of the surface as field evaporation occurs. Figure 4(c) shows a slice from the 3DFIM stack showing planes distorted by this simple reconstruction and, within the planes, surface atoms are clearly imaged. These images are similar to the 3DFIM image stacks reported previously (Klaes et al., 2021; Vurpillot et al., 2007). A surface atom produces a plume of field ionized imaging gas atoms that extends through $z_i$ until it evaporates. The plane interspacing is seen to be variable throughout the course of the experiment, perhaps due to the slow evolution of the specimen's radius, and also dependent upon the slowly decreasing detection rate over the course of the experiment which related in part to the decreasing helium pressure (admitted through the manual but unsupervised leak valve).

### 3.3. Multihit detection

We found that multiple hits in an aFIM experiment provided clearer elemental information than the unprocessed mass-to-charge spectrum (Katnagallu et al., 2019). Ions of a multiple event ($I_1$, $I_2$, $I_3$,..) can be subdivided as pairs $\{(I_1, I_2), (I_1, I_3), (I_2, I_3)\}$. In the past, experimental phenomena have been identified and analysed in multiple correlation plots (a.k.a. "Saxey - plot" (Saxey, 2011)) plotting the paired mass-to-charge ratios against each other. Multiple ions can be associated with each other in both time and space as previously reported (Müller et al., 2011; De Geuser et al., 2007). These ion pairs may not all be adequately discriminated as the paired hit distribution in its ($\Delta X$, $\Delta Y$, $\Delta t$) space-time coordinates demonstrates "dead zones" that are characteristic of the detector (Meisenkothen et al., 2015; Peng et al., 2018). Phenomena which can be identified in the correlation histograms are clusters of pulsed field evaporation, continuous "DC" field evaporation, and ion dissociation (Blum et al., 2016).

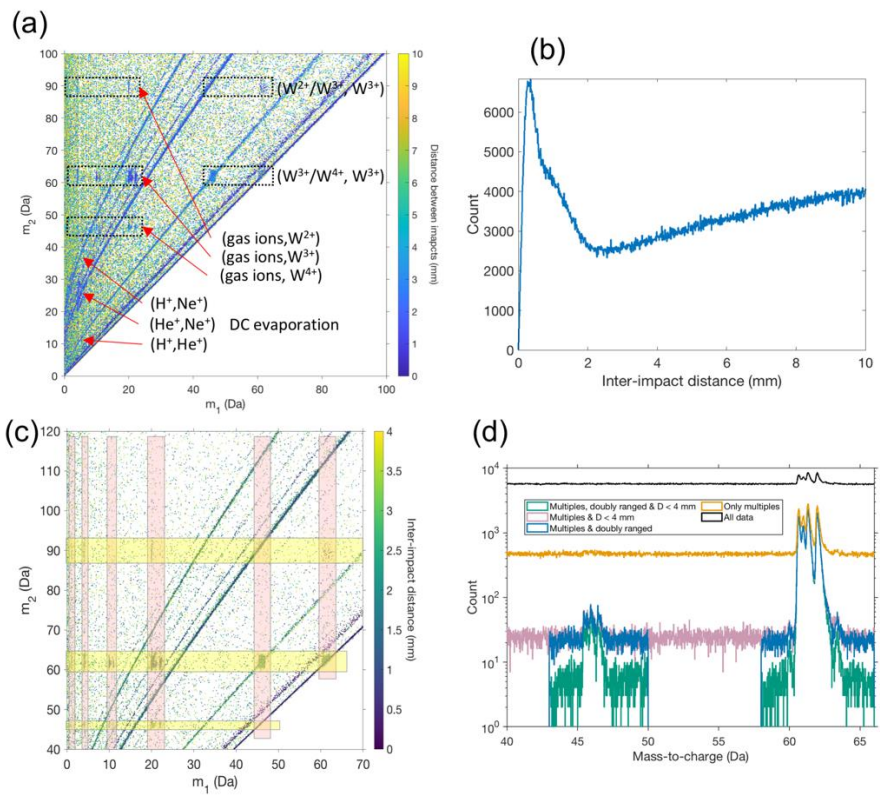

Figure 5. Multiple event detection and filtering by inter-impact distance and double ranging for the data featured in Figure 4. (a) Discerning mass-to-charge spectrum peaks through a correlation plot. The DC evaporation lines belong mainly to the gases ($\{H^+/Ne^+\}$ $\{He^+/Ne^+\}$ $\{H^+/He^+\}$) and the y=x line we did not identify as being exclusive to one species (this line corresponding to DC evaporation of ions with like mass-to-charge values). The boxes mark the co-evaporation of W with adsorbed gas species but also between $W^{3+}$ and $W^{4+}$ ions. (b) Distances between multiple ion impacts, displaying the correlation plot's colour information as a simple histogram which seems to have three components. (c) The concept of intersectional double ranging shown in relation to a correlation plot. (d) The total mass-to-charge spectrum (black) compared with the multiple spectrum (orange), and the inter-impact distance filtering (pink) and the double ranging (blue) done separately and together (green).

Figure 5 shows the step-by-step the mass-to-charge spectrum filtering process. Figure 5(a) shows the multiple correlation plot for the detected multiple events from a tungsten aFIM experiment. No ion dissociation was evident. This plot showed that both the $W^{3+}$ and $W^{4+}$ ions co-evaporated with the desorbing gas ions $Ne^{+/2+}$, $He^+$ and $H^+$ and that the tungsten co-evaporated with itself in $\{W^{3+/4+}, W^{3+}\}$ pairs, which can be associated to the rearrangement of charges subsequent to the departure of a first atom that causes the field evaporation of a second one (Katnagallu, Dagan, et al., 2018; De Geuser et al., 2007). There were also further pulsed co-evaporation events between $Ne^{+/2+}$, $He^+$ and $H^+$. We also observed a small continuous field evaporation signal, present as the positively-sloped diagonally-curving lines, identified as being mostly between desorbed $Ne^+$, $He^+$ and $H^+$ and not involving metal ions.

The colour scale described the impact separation distance that was measured between detected events. If this distance was large then the compared multiple events were likely physically independent. Conversely, if there were some physical associations, then it was likely that the inter-impact distance was much smaller than what you would expect randomly. This has been demonstrated before (De Geuser et al., 2007) and we have before exploited this to first give contrast to such events (Katnagallu et al., 2019). Figure 5(b) shows a histogram of the inter-impact distances of the many correlated ion pairs, condensed from Figure 5(a). At inter-impact distances greater than 4 mm, the large tail of this histogram was the dominant component representing the field ionisation background.

Ranging the field evaporated ions correctly was then a problem not of loss but one of error, meaning many ions *not tungsten* would be ranged *as* tungsten. Some mis-ranged ions came from the field ionisation signal, present in Figure 5(a) as the green-yellow background, suggesting that this signal could be filtered by a threshold on the inter-impact distance. Other mis-ranged ions were from the continuous field evaporation signals of paired gas ions. These signals had approximately the same inter-impact distances as the tungsten-tungsten co-evaporation clusters so an inter-distance threshold filter alone would not have worked to distinguish gas and metal. Figure 5(c) demonstrates an alternate filtering concept where we double-ranged upon multiple ion events taking only the ranging overlap area (taking out most of the continuous field evaporation background). Figure 5(d) shows the effect of applying these filters, separately and altogether. It can be seen that, not only did the $W^{3+}$ signal become clearer without significant loss of height but the mass-to-charge background was sufficiently reduced to make some $W^{4+}$ peaks conspicuous. The signal-to-background is quantified and discussed in section 4.3.

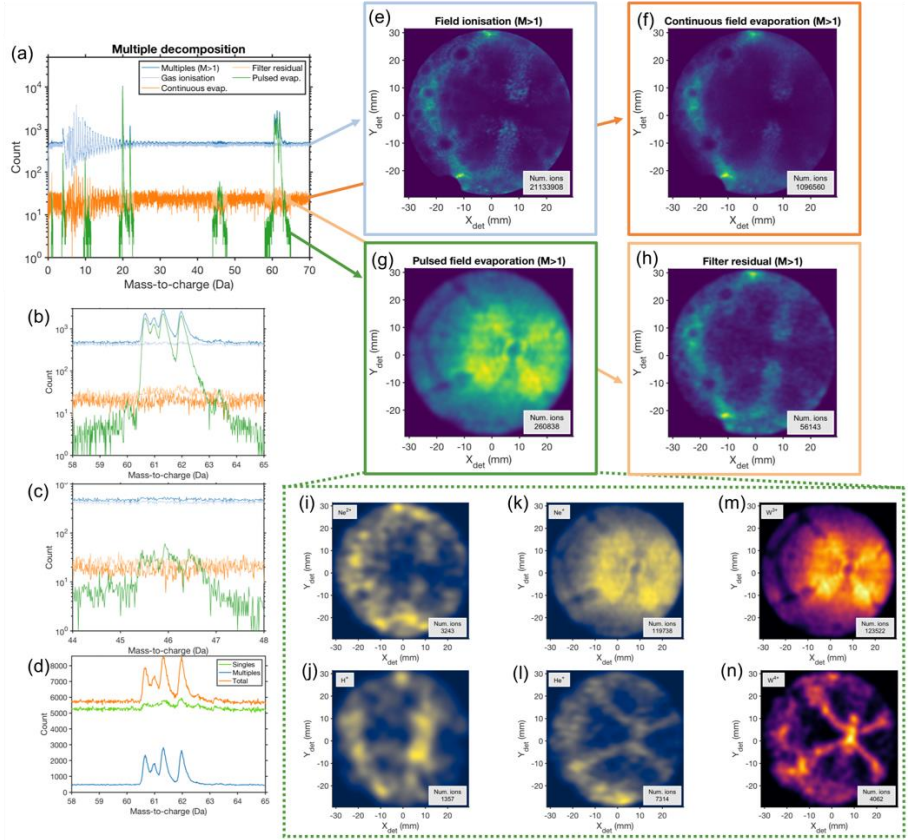

*Figure 6. The multiple detection events spectrally decomposed and classified as distinct physical components. (a) The multiple events (dark blue) can be split into an ionised gas signal (light blue), a continuous DC evaporation signal (dark orange), a pulsed evaporation signal (green) and the residual (light orange). (b) Inset of the $W^{3+}$ peaks with the same legend as (a). (c) Inset of the $W^{4+}$ peaks with the same legend as (a). (d) A simple decomposition of the total detected ions (orange) into singles (green) and multiples (blue) showed that most of the evaporated W signal came from multiple evaporation. Background subtraction suggests that there are approximately 127000 $W^{3+}$ ions in the multiples and approximately 40000 $W^{3+}$ in the singles spectrum. (e-n) The spectral decomposition and classification suggests different imaging modes. Specifically for the different physical components, we show here the detector maps for (e) field ionisation (f) continuous field evaporation (g) pulsed field evaporation and (h) the residual. The pulsed field evaporation was ranged and further broken down into (i) $Ne^{2+}$ desorption, (j) $H^+$ desorption (k) $Ne^+$ desorption, (l) $He^+$ desorption, (m) $W^{3+}$ evaporation and (n) $W^{4+}$ evaporation. For each of these maps, we used the 2D kernel density estimation implemented in Figs 2-4 but with a 1024x1024 pixel resolution.*
1212
12

Figure 6 offers an alternate spectral decomposition to Figure 5(d). Figure 6(a) shows that the multiple mass-to-charge spectrum can be broken down into field ionisation signal (light blue), continuous correlated field evaporation (orange), pulsed field evaporation (green) and the residual (light orange). Figures 6(b) and 6(c) show this same decomposition zoomed in upon the $W^{3+}$ and $W^{4+}$ peaks. Figure 6(d) demonstrates that most of the field evaporated signal for tungsten is actually contained in the multiply-detected events and not in the singly-detected events as is normal for atom probe tomography. Different imaging modes can be derived from the spectral decomposition as demonstrated by the detector maps shown in Figure 6(e) (field ionisation), Figure 6(f) (continuous field evaporation between H, He, Ne species), and Figure 6(g) (pulsed field evaporation of mostly W with itself and gases). Figure 6(h) shows the detector map of residual events which most resembles the field ionisation signal. However, there is some small suggestion in Fig 6(b) of unfiltered $W^{3+}$ which has been cast aside in this classification. The amount of ranged desorbed gas ions roughly matched that of the ranged $W^{3+}/W^{4+}$ ions. Figures 6(i-n) shows the ranged decomposition of the field evaporated events of Figure 6(g) for $H^+$, $Ne^{2+}$, $Ne^+$, $He^+$, $W^{3+}$ and $W^{4+}$ respectively. Similarities in the density distributions can be observed between the $Ne^+$ and $W^{3+}$ detector maps and the $He^+$ and $W^{4+}$ detector maps.

*3.4. Classification of surface atom FIM images*

We distilled information about the individual surface atom images from the FIM image stack of Figure 4(b). Instead of the Gaussian kernel peak finding method used in Figure 2(a), which required the selection of multiple hyperparameters, we used another 3rd party peak finding method called *extrema2* (Aguilera, 2021) which quickly and efficiently identified the majority of imaged surface atoms. This was demonstrated by Figures 7(a) and 7(b). The image's dynamic range was substantial, as shown by Figure 7(c), but effectively spanned 2-3 orders of magnitude with respect to the surface atom images' peak values.

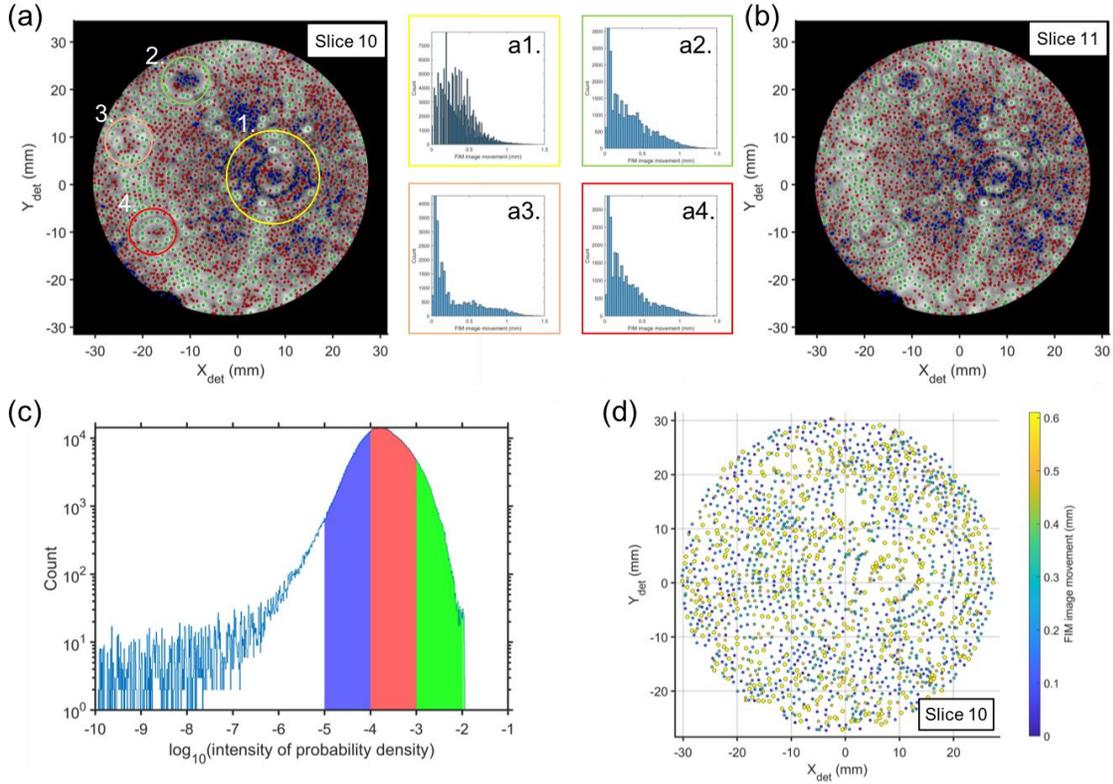

*Figure 7. A facile demonstration of a FIM image movement analysis classifying field evaporation. (a/b) The tenth and eleventh images of the 561-slice FIM image stack. The atoms' images are identified with the approximate extrema-finding algorithm (see text) and are each color-coded according to the relatively intensities at the local image maxima. (a1-a4) The 1st nearest neighbour distance histograms for the FIM image movements for the entire 561-image stack in the detector regions circled in Figure 7(a). (c) The histogram of the logarithm probability density values of the FIM image in Figure 7(a) showing the dynamic range of identified maxima values span over a few orders of magnitude, with coloured ranges corresponding to the extrema classification of Figures 7(a) and 7(b). (d) The red and green extrema in Figure 7(a) (probability densities $> 10^{-4}$) are here colour-coded by their subsequent movement in Figure 7(b), with atoms drawn larger if the movement exceeds 0.61 mm.*

Operating upon consecutive image slices as was done before (Klaes et al., 2021), we investigated the movement of surface atom images by calculating the 1$^{st}$ nearest neighbour distance measurements from the extrema positions of the current slice to the extrema positions of the subsequent slice. Taken over the entire 561-slice image stack, Figures 7(a1-a4) show the 1NN distance distributions for the corresponding regions marked in Figure 7(a). The averaged-over-time 1NN distance distributions are mostly continuous but Figure 7(a3) displays some discernible bimodality. Figure 7(d) displays a classification of the brightest images of Figure 7(a), where a 1NN distance threshold (d > 0.61 mm) was used to classify the surface atoms corresponding to the images as "evaporated" and its last measured detector coordinates stored. Here, this threshold was chosen simply because it resulted in the registered evaporated surface atoms (166154) approximately matching the detected tungsten ions divided by detector efficiency (127575/80% = 162500).

## 3.5. Partial synthesis of 3DFIM and atom probe tomography reconstructions

Figure 8 presents complementary and simultaneous reconstructions corresponding to the parallel imaging modes of field ion microscopy and atom probe tomography. The data used was the same as for Figures 3-7. Figure 8(a) displays a 3DFIM reconstruction made from information distilled from the image stack (Figure 7). Figure 8(b) displays the corresponding APT reconstruction including only the filtered and ranged $W^{3+}$ and $W^{4+}$ ions (Figure 6).

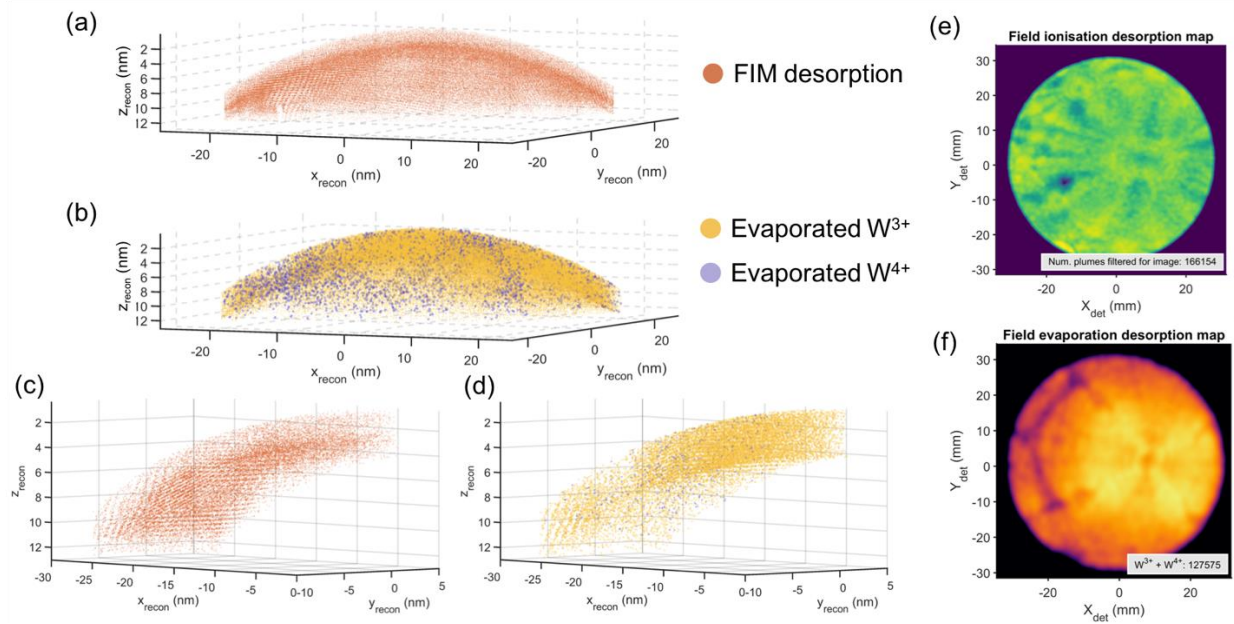

*Figure 8. Parallel 3DFIM and APT reconstructions using an azimuthal equidistant projection (k = 0.21) and modelling a hemispherical (r = 46 nm). (a) The 3DFIM reconstruction with every point corresponding to a single registered evaporation. (b) The APT reconstruction with every point corresponding to either $W^{3+}$ (dark yellow) or $W^{4+}$ (light blue). (c/d) Magnified region near [121] poles for the 3DFIM and APT reconstructions respectively. (e/f) 1024x1024 kernel density estimates representing (e) the FIM projection of all 16654 registered evaporations and (f) the field evaporation projection of all 127575 filtered and ranged tungsten ion impacts.*

A reconstruction protocol using an azimuthal equidistant (AE) projection was employed (De Geuser & Gault, 2017). Between the [022] and [121] poles, we measured k = 0.21 in the AE projection relationship $\theta = kD_{detector}$. No care was taken here to include the bowl centre in the projection measurement. Ring counting between the [022] and the [031] poles suggested a local radius of 46 nm (Drechsler & Wolf, 1958). For the purposes of demonstration, both the k-value and the radius were taken as global reconstruction parameters. Figure 4(c) showed that the evaporation rate of planes fluctuated over the course of the experiment (Chen & Seidman, 1971).

Figures 8(c) and 8(d) show a magnified region around the [121] pole (at $X_D, Y_D$ = [-19, -11]) for the 3DFIM and APT reconstructions respectively. The 3DFIM reconstruction clearly show intersecting planes corresponding to the central [022] pole (which was visible throughout the reconstruction's entire breadth) and the [121] pole. These planes are also visible in the APT reconstruction but only in the vicinity where the planes are tangential with the specimen surface. A third plane is clearly seen in the 3DFIM reconstruction corresponding to an unidentified pole outside of the available field-of-view, but was probably [220]. Surprisingly, this [220] plane is also visible in the APT reconstruction. In other respects however, the capability of the field evaporated signal is as for APT, where multiple planes are visible but are not present in the same volume.

The convergence of the three planes in the 3DFIM reconstruction suggested that atomic resolution was achieved. However, Figure 8(e) and (f) show the different projections for the FIM signal and the field evaporation signal, showing that the two reconstructions are spatially different (Krishnaswamy et al., 1975; Waugh et al., 1976; Klaes et al., 2021).

## 4. Discussion

We have here reported our attempt to consolidate the well-established techniques of (3D)FIM and APT, each providing an advantage that the other lacks. Culminating our work-to-date in Figure 8, we can confidently say that we have provided proof-of-concept for analytical FIM (aFIM). We highlighted one way how this may be done, and we shall discuss our particular steps in detail below, suggesting where improvements may be required.

*4.1. Data acquisition*

We acquired our data upon our Cameca LEAP 5000XS and LEAP 3000HR using Cameca's proprietary software (DaVis & LCC) to which there is no alternative. We could exert little control upon the instrument or upon the acquisition process beyond what was normally available for atom probe experimentation. Though worthwhile, an on-the-fly analysis was not programmable and so there could be no control feedback. This is on one small hand good, suggesting that our experiments can be easily replicated by other groups, but on the other hand, leaves much to be desired given that there are various operations and measurements that are inaccessible. We here discuss the simple things that we performed in spite of these limitations.

Figure 2 showed that a 2D kernel density estimation of ion impact coordinates provided a reasonably clean FIM image via probability density estimation. We did not canvass all KDE techniques and only employed an established 2D-KDE-by-diffusion method which seemed to serve well (Botev et al., 2010). We instead focused upon the effect of the experiment's acquisition parameters which were observed to introduce significant variability in the image quality and its information content.

One challenge was to optimise APT parameters to produce reasonable FIM images while also simultaneously field evaporating a few atomic layers. We explored this compromise in the FIM image sequence of Figures 3(c/e/g/i). With the standing field held constant, the pulse fraction was increased and this increased the field evaporation rate, changing in turn the imaging conditions. The information that a FIM image contains is thus dependent upon exposure time and the field ionization flux to give a particular field ionization dose (equals exposure time multiplied but field ionization flux). However, because field evaporation causes surface change, image quality also depends upon the ratio between field evaporation and field ionization fluxes. A smaller ratio means a cleaner image because there will be less field evaporation and less surface change averaged during the exposure. This was clear in the transition from Figure 3(e) to 3(g) to 3(i). As demonstrated by the two parts of Figure 3(i), a shorter exposure can compensate for an increased evaporation rate. The specific image structure and its dynamic range will also affect the ability to discriminate between neighbouring surface atoms.

*4.2. Image analysis*

FIM naturally yields image data, whether "eFIM" or from a camera viewing a phosphor screen, and slow surface field evaporation can produce videographic data that may be used for 3DFIM reconstructions (Vurpillot et al., 2017). In contrast, we started from the quantized ion detection data typically acquired for APT. For imaging, we considered only the ions' detector coordinates and their sequence number. We then constructed 2D-KDE FIM images by slicing the detector+sequence dataspace on the basis of sequence number (using this as a proxy for time). This transformation from point data to FIM imaging is a novel intermediate step which provides a finely controlled image construction. However, it is presently only something that can be done post-analysis with the LEAP systems which can make it a great computational burden, not only in terms of data storage but also of course in memory management and running time when loading and analyzing data. Figure 4(a) was comprised of approximately 280 million ions and from that we constructed 561 exposures to make Figure 4(b). Through the filtering encapsulated by Figure 6, it was shown that very few of the total detected ions were actually tungsten, the rest mostly being from the helium and neon field ionization. We estimated that there were between 170 and 180 thousand field evaporated tungsten ions detected (considering the singly detected ions and accounting for background level) meaning that the materials' field evaporation was only about 0.06-0.07% of the total data. This low yield is directly related to the evaporation-ionization flux ratio and, as discussed above, a low value is likely necessary to generate sufficient FIM image quality. Despite being excruciatingly slow when compared to APT, this slow evaporation flux would help to preserve the sample as higher values of the detection rate parameter has been attributed to be the higher occurrences of sample fracture.

Within each exposure, we identified individual surface atom images using an approximate extrema-finding algorithm (Aguilera, 2021). This algorithm was selected over convolutional peak finding (Dagan et al., 2017; Katnagallu, Gault, et al., 2018) for its quick and adequate functionality. However, not every surface atom image should be registered as an evaporated ion, and we departed from recent studies which identified evaporation by significant non-zero signals in the subtraction of two consecutive images (Klaes et al., 2021). We adopted a similar outlook but considering our earlier work on FIM imaging instead which discussed how a surface atom's image suffered little displacement prior to evaporation (Katnagallu, Dagan, et al., 2018). Based on this finding, evaporation registration was marked by any significant distance shift between a slice's extrema locations and the extrema locations of the subsequent slice. This "significant distance" (d = 0.61 mm) here was a threshold chosen heuristically to match the population of the field evaporation signal found in Figure 6. Likely, a union of the two techniques could produce more reliable identification and improvements are required upon the FIM imaging in an aFIM experiment. Figure 7(a) and the inset Figure 7(a3) suggested that higher indexed poles may give better discrimination with respect to image movement and so a sub-volume may be more easily studied that the global scope of an entire experiment. This has been earlier demonstrated in our 3DFIM study on Ni-Re (Katnagallu et al., 2019).

*4.3. Mass-to-charge spectrum analysis*

In early APFIM, the time-of-flight mass spectrometer tube was differentially pumped with respect to the analysis chamber (held at $10^{-3}$ Torr image gas pressure). The image gas ion current would amount to $10^3$ to $10^4$ ions per second per imaged surface atom (Mclane et al., 1969; Müller et al., 1968). This was only a minor issue for APFIM when, with its small probe hole sampling, an energy-deficit compensation system like a Poschenrieder lens could be used (Sakai & Sakurai, 1984). But this was obviously going to be an issue for our experiments where the field evaporation to field ionisation ratio needs to be high enough to resolve peaks in the mass-to-charge spectrum but low enough to provide the required high dynamic range of a good FIM image.

As expected, time-of-flight mass spectrometry gave nothing useful without a significant pulse as shown by Figure 2(b). Field ionised events exhibited an approximately uniform time-of-flight spectrum. Even with relatively high backgrounds, field desorption and field evaporation peaks were evident when higher voltage pulsing was introduced as demonstrated in the spectra of Figure 3. Another issue was that the massive field ionisation background causes the commercial software to fail in making the necessary time-of-flight corrections. We reported only the time-of-flight spectra in Figures 3(d) and 3(f) for this reason. For the other spectra (Figures 3(h), 3(j) and 5(d)), enough field evaporated signal was present for the commercial software to attempt the required time-of-flight corrections. Once first-pass filtering has been accomplished, this can better be performed according to established protocols (Sebastian et al., 2001).

Analysis performed here upon the time-of-flight spectrometry relied upon the filtering of a large fraction of multiple detection events enabled by the physical correlations between them. Such enhancement has been demonstrated before (Yao et al., 2010; Katnagallu et al., 2019). This can be explained by gas atoms forming an adsorbed surface layer (Wang et al., 1996), and it has been proposed that a field evaporation event can act as a catalyst for gas field desorption (Mclane et al., 1969). We implemented filters based upon correlated evaporation events (double mass-to-charge ranging – Figure 5(c)) and the close physical situation such events imply (relating a close impact distance to their initial close proximity on the surface – Figure 5(b)). We applied these filters on the global data, but they may be better applied on a local basis in order to further remove background and identify more field evaporated signal.

Additionally, the mass-to-charge spectra revealed that, although the helium inserted was thought to be 99.999% pure, contamination of Ne-in-He was evident in our aFIM experiments and perhaps this was more pronounced owing to a neon atom's higher polarizability and thus higher attraction to the high-field specimen tip. We took advantage of both the neon gas adsorption and its catalytic field desorption, and the main association that enabled this approach in this manuscript were between tungsten and neon ions. Other co-evaporation events (W/W, W/He or W/H) appeared as well. For example, the signal pertaining to correlated tungsten-hydrogen evaporation in Figure 5(a) was low, but the corresponding inter-impact distance for this co-evaporation was larger than for tungsten-neon or tungsten-tungsten evaporation. This may be related to the mass of the ion and the effects of dynamic effects from the pulse (Rousseau et al., 2020). Other metrics may provide different filtering options, not only upon multiple detection events, but also ions detected on successive pulses for instance (De Geuser et al., 2007).

What little background near the tungsten peaks that remained was likely drawn from the continuous field desorption and field ionization of gas ions that distribute into those ranges even after double ranging and filtering upon the inter-impact distance. Extrapolating the cumulative background signal prior to the $W^{3+}$ peaks (to an equation $at^2+bt+c$ where $t^2 = m$), we calculated the signal-to-background ratio (SBR) for the $W^{3+}$ peaks to be 0.062 without any filtering, 0.61 retaining only multiple events, and 66 concerning only the filtered pulsed evaporation $W^{3+}$ of Figure 6(b) and 6(m). This means that 1/67 ranged $W^{3+}$ ions in the filtered pulsed evaporation may not be tungsten. Achieving this 1000-fold improvement alone was of merit and deserves further investigations but also further refinements because we calculated a high value of 832 $W^{3+}$ SBR for the preceding APT data set in Figure 3(b).

Chemical identification of field evaporated ions can perhaps be improved by examining not only multiple correlations but the point processes behind field ionization. The emissions or ion plumes are clearly seen in Figure 4(c) but their original form is a cloud of points contained in the point cloud of Figure 4(a). In an image stack like the one presented in Figure 4(b), previous approaches have considered differences in pixel-to-pixel intensity between two images (Klaes et al., 2021). Given that multi-modal aFIM data, a more straight-forward approach would be to consider the data structure in the more primitive 3D point cloud.

*4.4. Hardware considerations*

Limited detection efficiency is as much as issue for aFIM as it is for all APT experiments, and some information is lost. Another concern was the detector's capacity to identify individual hits from the delay line detector (Jagutzki et al., 2002; Da Costa et al., 2005; Costa et al., 2012). Despite the large flux compared to APT experiments, this did not manifest much in the experiments reported herein, except in a few occasions when the commercial software reported that the detection rate exceeded what was permitted and some events were not recorded. With respect to the imaging gas, it was found that the solution packaged with the LEAP5000XS was adequate for standard FIM imaging but needed improvement for controlled aFIM experiments. Automatic gas mixing, gas analysis via quadrupole mass spectroscopy, and gas insertion via a computer-controlled leak valve are three developments which could assist in providing the necessary experimental rigour. Experimental control also requires fine-tuned maintenance upon the LEAP5000XS.

*4.5. Ripples in mass-to-charge spectrum*

The damped artefact ripples which were observed in the spectra of Figure 2(b), Figures 3(d/f/h/j) and Figure 6(a). One common aspect was that these features followed the imaging gas peaks which, despite the apparent neon contamination that helped filtering, was helium for the tungsten example used in the body of the manuscript. To elucidate how they formed, we obtained aFIM data on two instruments to isolate any instrumental effects.

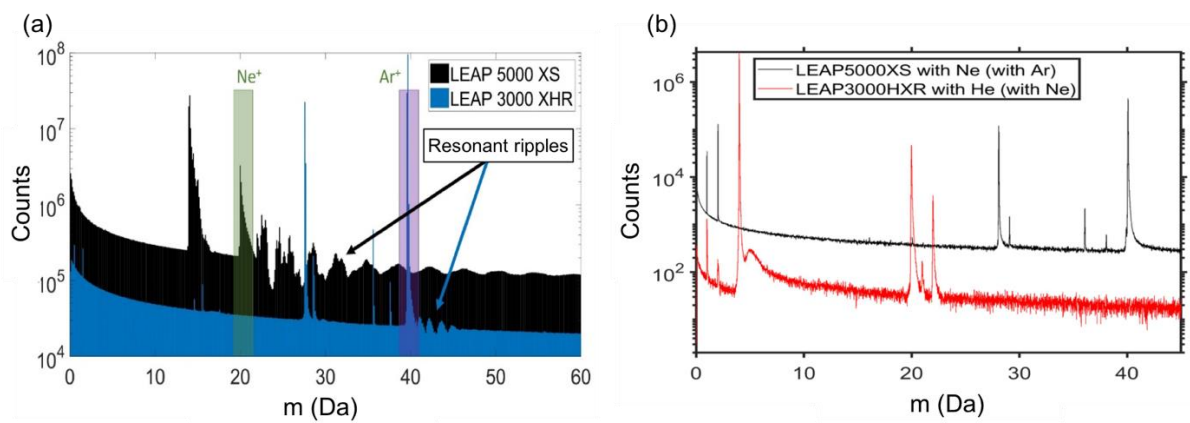

*Figure 9: (a) Two aFIM experiments upon pre-sharpened silicon microtips using the LEAP5000XS with neon imaging gas and the LEAP3000XHR with argon imaging gas. "Resonant" peaks are seen in both spectra following the image*

*gas peaks. (b) No resonant peaks were observed during the laser pulsing experiments for two aFIM experiments upon tungsten using the LEAP5000XS with Ne imaging gas (with significant presence of adsorbed argon) and the LEAP3000HXR with helium imaging gas (with significant presence of adsorbed neon).*

Two different voltage-pulsed aFIM datasets were obtained on commercial, pre-sharpened silicon microtips (Cameca Instruments, Inc.). One using neon imaging gas on the LEAP5000XS and one using argon on the LEAP3000XHR. Like for the experiment figuring in the main body of the manuscript, the standing voltage was held constant. Figure 9(a) showed that the mass-to-charge spectra obtained from these experiments both exhibited ripples after neon (20 Da) and argon (40 Da) respectively. For the latter experiment on the LEAP3000HR, the presence of the ripples demonstrated that this was not a result of an energy deficit that can be remedied with a reflectron. The ripples were associated with the imaging gas and their effect was manifest directly after the imaging gas peaks, so we naturally deduced that they were some interaction with the imaging gas ions. One hypothesis was that it was induced by the voltage pulser. Supporting this, Figure 9(b) showed two additional laser pulsing experiments on tungsten, the first on the LEAP5000XS using neon imaging gas and the second on the LEAP3000XHR using helium imaging gas, and neither exhibited the artefact ripples.

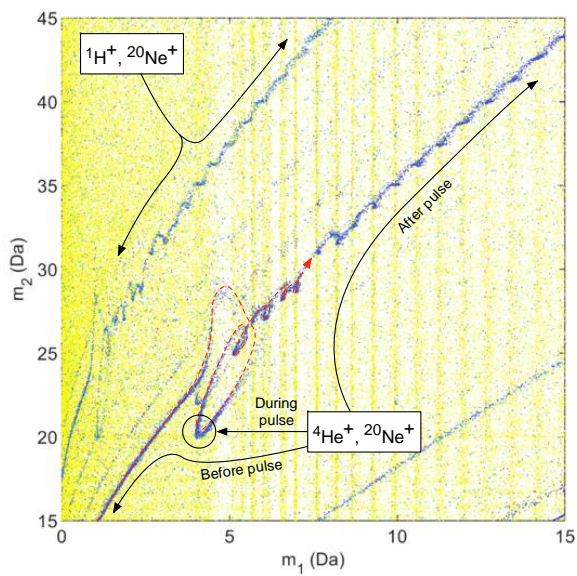

*Figure 10. The two most intense DC evaporation lines in the multiple correlation space here corresponding to the co-evaporation of $H^+$ and $Ne^+$ (top) and $He^+$ and $Ne^+$ (dark band below that). The red dashed line partially traces the correlated continuous field evaporation line for the co-evaporating $Ne^+$ and $He^+$ which, during the peak of the pulse, passes through the (4,20) coordinate corresponding to the ($He^+$, $Ne^+$) pair.*

The ripples were also somewhat evident in the multiple correlation plot of Figure 5(a) and Figure 10 shows a magnified view of them. Seen nearby were the two correlated continuous field evaporation (CCFE) lines in the multiple correlation space. These lines were associated with the ($m_1,m_2$) mass-to-charge coordinates they pass near when the timing of the paired evaporation coincided with the $t_0$ pulse time. Ideally, they should pass through the ($m_1,m_2$) coordinates but are slightly higher than they should be and this is illustrated in Figure 10. The CCFE lines corresponded to the co-evaporation of ($H^+/Ne^+$) and ($He^+/Ne^+$). Continuous evaporation occurred at the standing voltage (held constant throughout these experiments) and indeed, at constant voltage, the CCFE lines are tracers which have a constant parameterisation by time-at-evaporation relative to the voltage pulse.

During the pulse, the ions upon this curve were virtually identical with pulsed field evaporation ions, evaporate with the same energy and were measured with a correct time-of-flight which gives the correct mass-to-charge. This was seen – there was one cluster at (1,20) representing the ($H^+$/$Ne^+$) co-evaporation and another cluster at (4,20) representing the ($He^+$/$Ne^+$) co-evaporation. The artefact and its phenomena at (4,20), has been explained recently (Rousseau et al., 2020) and the ripples can be thought of as a continuation. Away from these points, the curving lines passing through these points are generally shifted to larger masses because they evaporate at a lower voltage and so attain less energy. But near the co-evaporation points the lines were wildly distorted and oscillations in this line persisted over a large range, matching up with the yellow bands that correspond to the 1D mass-to-charge spectrum peak artefacts. The curving CCFE lines also suggest that the pulse subtracts from the standing voltage right before the pulse peak.

Impedance mismatching of the voltage pulsing line likely causes fluctuation in the bias upon the specimen tip and field ionisation, field desorption and field evaporation are all both affected similarly by this artefact. This artefact will also be present in our LEAP 5000XS machine where non-correlated evaporation is taking place and it would occur in a voltage-pulsed APT experiment. In contrast to normal APT operation, they were evident in our experiments because of the abundance of light imaging gases (Rousseau et al., 2020) and because we kept the standing voltage constant, prevented this artefact being smeared. One obvious solution would be to employ laser-pulsing in our future aFIM experiments until this issue can be resolved.

*4.6. Perspectives*

Future aFIM investigations will be applied to materials that are amenable to FIM imaging, i.e. alloys with ordering, clusters and decorated interfaces. There remains substantial work, mainly in software development, but also in revisiting much of the high field surface science physics underpinning field emission microscopies. Realizing aFIM data that is a complete synthesis of APT and 3DFIM data, as opposed to the partial synthesis presented here, will be a task requiring multifarious improvements at all stages of the experiment. Broadly, we can say that acquisition, data analysis and reconstruction must be improved, and maybe more tightly incorporated together.

An aspect is the control and the analysis software, especially for commercial atom probes. An aFIM experiment needs to be understood as it is happening for direct experimental feedback, and this is currently not possible. A reasonably-sized aFIM dataset could be in excess of 100 gigabytes and an analysis feature space could explode the memory requirements well above the terabyte level, and a sizeable fraction of this data may not be of use. Full analysis of aFIM experiments is well beyond the computational capabilities of many institutions and would besides be expensive in time and resources. Only a fully realizable online approach, requiring sophisticated instrumental control and a comprehensive on-the-fly analysis such that no software currently provides, will enable the development of an accurate and revealing chemically-labelled 3DFIM. It is the belief of the authors that the necessary hardware (commercial or otherwise) already exists in many atom probe labs around the world. For the commercial instruments, scientific advancement requires collaboration and negotiation between industry and academia. With regards to the analysis and reconstruction packages, they are often assuming clean APT experiments and are in any case still reliant upon commercial software to parse the data into an appropriate format. None of this software gives adequate control over the microscope and so all of this software is wholly unsuited for this aFIM data synthesis. This is because it was designed and optimized for typical APT data. Data for aFIM is larger and can produce at least 1000x more raw data for the same probed volume, and can possibly contain more information about surface the atoms prior to field evaporation.

## 5. Conclusions

To summarise, we have shown that it was possible to simultaneously acquire dynamic field-ion images and pulsed time-of-flight spectrometry concomitantly. We demonstrate this using the voltage-pulsed mode in the analysis of pure tungsten. We decomposed the TOF spectrometry and applied filters, leaving little residual unassigned. Some tungsten was evidenced in the singly-detected spectrum albeit with a low signal-to-background. The signal-to-background for the filtered $W^{3+}$ peak was 1000-times better than for the unfiltered spectrum signal (0.062) but approximately 10-times worse than for a comparable APT experiment (832). We produced a 3DFIM reconstruction by constructing smoothed FIM images from the {detector coordinates, ion sequence} dataspace, identifying all individual ions in these FIM images, marking evaporated ions by significant lateral image displacements, and recording their last projected image position which are then used for a back-projection reconstruction. We then produced a matching APT reconstruction using events from the filtered TOF spectrometry from the same probed volume. We scaled the depth increments for each reconstruction differently so that they were of identical dimensions. For both, we employed a standard reconstruction protocols but with an azimuthal equidistant projection. These parallel and complementary reconstructions demonstrate a partial synthesis of two previously-independent imaging techniques. This new hybrid technique could soon allow for true atomic-scale imaging with analytical capabilities, and we have here established the motivation for further investigation.

**Acknowledgments**

S.K., L.T.S., I.M. and B.G. are grateful to Dr Richard Forbes for discussions made in mid-2018. S.K. and F.F.M. acknowledges financial support from the International Max Planck Research School for Interface Controlled Materials for Energy Conversion (IMPRS-SurMat).. L.T.S. acknowledges his pre-2017 association with Dr. Peter V. Liddicoat, notably the time spent in the start-up enterprise "Åtomnaut" in the University of Sydney's 2015 "Incubate" accelerator. L.T.S. and B.G. acknowledge financial support from the ERC-CoG-SHINE-771602. The authors are grateful to U. Tezins and A. Sturm for their technical support of the APT at the Max-Planck-Institut für Eisenforschung.